%Paper: hep-th/9211020
%From: gamboa@ipncls.in2p3.fr
%Date: Wed, 4 Nov 92 13:19:09 GMT

\magnification=1200
\footline={\ifnum\pageno=1\hfill\else\hfill\rm\folio\hfill\fi}
\font\title=cmr12 at 12pt
\baselineskip=18pt
\def\dsl{\raise.15ex\hbox{/}\kern-.57em\partial}
\def\Dsl{\,\raise.15ex\hbox{/}\mkern-13.5mu D} %this one can be subscripted
\def\Asl{\,\raise.15ex\hbox{/}\mkern-13.5mu A} %this one can be subscripted
\def\Bsl{\,\raise.15ex\hbox{/}\mkern-13.5mu B} %this one can be subscripted
\vskip 1.0cm
\centerline{{\title HAMILTONIAN APPROACH TO }}
\vskip 0.1cm
\centerline{\title 2D SUPERGRAVITY}
\vskip 1.7cm
\centerline{\bf J. GAMBOA}
\vskip 0.05cm
\centerline{\it Division de Physique Th\'eorique,}
\centerline{{\it  Institut de Physique
Nucleaire}
\footnote{$^1$}{\it Unit\'e de Recherche des Universit\'es Paris 11
et Paris 6 associ\'ee au CNRS.},}
\centerline{\it Orsay 91406, France.}
\vskip 0.15cm
\centerline{{\bf C. RAMIREZ}}
\centerline{\it Facultad de Cs. F\'{\i}sico Matem\'{a}ticas}
\centerline{\it Universidad Aut\'onoma de Puebla}
\centerline{\it A.P. 1364, Puebla 72000, M\'exico.}
\vskip 2.5cm

{\bf Abstract}. We study $2D$ supergravity in a covariant and gauge independent
 way. The
theory is obtained from $2D$ bosonic gravity following the square root
method and the diffeomorphism superalgebra is explicitly computed.
We argue that our approach could be a procedure for introducing nontrivial
physics in quantum $2D$ (super)gravity.
\vskip 1.5cm
\leftline{IPNO/TH 92/79 }
\leftline{CFUAP-92-10}
\leftline{October 1992}
\vfill
\eject

Some years ago Polyakov observed that string theory can be understood as $2D$
 quantum gravity coupled to $d$ scalar fields {\bf [1]}. When $d=26$,
the scalar
fields decouple from gravity and the quantization of strings can be
made without understanding the quantization of $2D$ gravity.

The problem of quantizing $2D$ gravity has been considered
by Polyakov {\bf [2]},
by Knizhnik, Polyakov and Zamolodchikov (KPZ) in the light cone gauge
{\bf [3]} and by David, Distler and Kawai
(DDK) {\bf [4]} in the conformal gauge.

The results reached by these authors are in complete agreement with numerical
simulations {\bf [5]} and with the matrix model approach {\bf [6]}.
However, in spite of the relevance that these results have in our
understanding of $2D$ quantum gravity, several problems still
remain.

One of these problems is the lack of a gauge covariant formulation for
quantum $2D$ (super)gravity that would permit us to understand in a
gauge independent way the results of KPZ and DDK.

Inspired in Marnelius work {\bf [9]}, this problematic has been
tackled recently for the bosonic theory {\bf [7-8]} following
the traditional metric formalism. However, it turns out that
this formalism is too involved to allow a transparent formulation
of Hamiltonian formalism, in {\bf [9]} the conformal gauge has
been eventually imposed in order to make things easier to handle.

The purpose of this note is, starting from the action of $2D$
gravity given by
$${\cal S} = -{1\over 2}\int{d^2 \sigma ( \sqrt{-g} g^{m n}
\partial_m \varphi \partial_n \varphi + {Q\over 2} R\varphi)}
\eqno(1)$$
to give a manifestly covariant hamiltonian formulation based
on the zweibein formalism from which then, by means of the
square root method, $2D$ supergravity is obtained.
In {\b [1]} $\varphi$ is a scalar field and $Q$ is a numerical coefficient.

Thus, we start considering the zweibein formulation of {\bf [1]}.
In the metric formulation the complication arises from the
involved form of the scalar
curvature in terms of the metric tensor. However, in terms of zweibein
formalism, the scalar curvature has a quite a simple form
$$R = \epsilon^{mn}\partial_n \omega_m, \eqno(2)$$
where
$$\omega_m = \varepsilon^{kl}\partial_l e_k^a e_{ma},\eqno(3)$$
is the induced Lorentz connexion, $e_m^a$ the zweibein and $\varepsilon^{kl}
= e \epsilon_{kl}$ the Levi-Civita invariant density, $e = \det (e_m^a)$ and
$\epsilon_{01}=1$.

In this way the variation of the action gives the equation of motion of the
scalar $\varphi$:
$$\partial_m (eg^{mn} \partial_n \varphi ) = -{Q\over 2}eR, \eqno(4)$$
and the constraints
$$\eqalignno{ &\Phi_a = 2\epsilon_a^b P^{\prime}_b + {4\over Q}P_\varphi P_a +
{4\over Q^2}  e_{a} P^a P_a =0, \cr &
\Phi = \epsilon^{ab} e_a P_b + {Q\over 2}\varphi^{\prime} = 0 &(5) \cr}$$
where $e^a = e_1^a$,
$P_a = {\delta L\over \delta {\dot e_1^a}}$, and $P_\varphi =
{\delta L\over \delta {\dot \varphi}}$,
$e_0^a$ being Lagrange multipliers.

The hamiltonian is obtained in the usual way and is given by
$$H = \int dx \biggl[ -{1\over 2}e_0^a \epsilon_a^b\Phi_b +
{(e_0^aP_a)}^\prime \biggr], \eqno(6)$$
where the surface term is similar to the four dimensional case and
can be discarded with suitable boundary conditions
\footnote{$^1$} {For a discussion on the modifications in the non compact
case see {\bf [10]}.}.

The constraint algebra closes ad is given by
$$\eqalignno{&[\Phi(x), \Phi(x^\prime)] = 0,
\cr & [\Phi(x), \Phi_a(x^\prime)] = \epsilon_a^b \Phi_b (x) \delta (x -
x^\prime), &(7) \cr}$$
and
$$[\Phi_a(x), \Phi_b(x^\prime)] = -{8\over Q^2} \epsilon_{ab}\epsilon^{cd}
P_c (x) \Phi_d (x) \delta (x - x^\prime). \eqno(8)$$

The constraint $\Phi$ describes Lorentz transformations with an induced
connexion ${Q\over 2}P_\varphi$.

The structure of the constraints $\Phi_a$ and of their algebra is not
very suggestive. Fortunately, after an appropiate canonical transformation on
$P_a$ and $e_a$
$$\eqalignno{ &P_a \to {\tilde P}_a  =  \bigg( -e^aP_a, \epsilon^{ab}e_aP_b
\biggr) \cr &   e_a \to {\tilde e}_a = \biggr(
{1\over 2}ln(e^a e_a),  {1\over 2} ln({e_1 - e_0 \over e_1 + e_0}) \biggr) &(9)
\cr}$$ and
$$\Phi_a  \to {\tilde \Phi}_a = \biggl( - e^a \Phi_a , \epsilon^{ab}e_a \Phi_b
\biggr), \eqno(10)$$
the structure of the constraints, as well as their algebra, simplifies
significantly
$$\eqalignno{&{\tilde \Phi}_0 = -2{\tilde P}^\prime_1 +2 {\tilde e}^\prime_0
{\tilde P}_1 -2 {\tilde e}^\prime_1
{\tilde P}_0 + {4\over Q} P_\varphi {\tilde P}_0 + {4\over Q^2}{\tilde P}^2
\cr & {\tilde \Phi}_1 =
-2{\tilde P}^\prime_0 + {\tilde e}^\prime_0
{\tilde P}_0 - {\tilde e}^\prime_1
{\tilde P}_1 + {4\over Q} P_\varphi {\tilde P}_1. &(11) \cr}$$

The algebra is given by the reparametrizations of the circle
$$\eqalignno{ &[\Phi_0(x), \Phi_0(x^\prime)] = ( \Phi_1(x) + \Phi_1 (x^\prime)
) \delta^\prime (x - x^\prime ),
\cr & [\Phi_1(x), \Phi_1(x^\prime)] = ( \Phi_1(x) + \Phi_1 (x^\prime)
) \delta^\prime (x - x^\prime ),
\cr & [\Phi_0(x), \Phi_1(x^\prime)] = (\Phi_0 (x) + \Phi_0(x^\prime))
\delta^\prime (x - x^\prime ), &(12) \cr}$$
as well as Lorentz invariance
$$ [\Phi (x), \Phi_b(x^\prime)] = 0. \eqno(13)$$

{}From (9) it is obvious that the constraints $\Phi$ decouple
(this similar to the constraint $\pi^0$ in gauge theories).
In light cone notation the remaining constraints can be written as
$$L_{\pm} ={1\over 2} (\Phi_0 \pm \Phi_1) =
 \biggl( h^2_{\pm} \pm {Q\over {\sqrt 2}} h^{\prime}_\pm - J^2_\mp +
{Q\over {\sqrt 2}} J^\prime_\mp \biggr), \eqno(14)$$
where
$$\eqalignno{& h_\pm = {\sqrt 2\over Q}{\tilde P}_0 -
{1\over{\sqrt 2}} P_\varphi +
{Q\over {\sqrt 2}}( {\tilde e}^\prime_1
 \mp {\tilde e}^\prime_0) \cr &
J_\pm = {\sqrt 2\over Q}{\tilde P}_1 \mp
{1\over{\sqrt 2}} P_\varphi \mp { Q\over {\sqrt 2}}( {\tilde
e}^\prime_1
 \mp {\tilde e}^\prime_0), &(15) \cr}$$
satisfy the algebra ($a,b=\pm 1$):
$$\lbrack h_a(x) , h_b(x^\prime) \rbrack =  2a \delta_{ab}\delta^\prime (x -
x^\prime)$$
$$\lbrack J_a(x) , J_b(x^\prime) \rbrack =- 2a \delta_{ab}\delta^\prime (x -
x^\prime) \eqno(16)$$

The last expressions correspond to left and right moving
sectors and the constraint algebra (11) is now
$$ \lbrack L_a(x) , L_b(x^\prime) \rbrack =
 a \delta_{ab} ( L_a(x) + L_a(x^\prime) ) \delta^\prime (x - x^\prime).
\eqno(17)$$

In fact, these results reflect the conformal invariance of the
theory and are similar to the ones obtained
from a gauge covariant treatment of string theory where the
constraints are:
$$T_{++}={{\sqrt{-g}}\over{4 g_{11}^2}}(g_{00}g_{11}-
2g_{01}^2) (P^2+{X'}^2)-{{gg_{01}}\over{g_{11}^2}} PX'$$
$$T_{--}=-{{\sqrt{-g}g_{01}}\over{4 g_{11}}} (P^2+{X'}^2)+
g{1\over{2 g_{11}}} PX',$$
which are equivalent, of course, to the usual quadratic constraints
$P^2+{X'}^2$ and $PX'$.

Having these results in mind, we construct $2D$ supergravity theory
following the square root method {\bf [11]}. In the context of four
dimensional supergravity theory it was first applied in {\bf [12]}.

We start making the following ansatz for the fermionic generators
$${\cal S}_a = \alpha \Gamma_a h_a + \beta Q \Gamma^\prime_a + \gamma
\Theta_{-a}  J_{-a} + \delta Q \Theta^\prime_{-a}, \eqno(18)$$
Awhere $\alpha, \beta, \gamma$ and $\delta$ are unknown contants and $Q$ is the
coefficient that appears in (1); here $\Gamma_a$ and $\Theta_a$ are two
dimensional real spinors that obey the Clifford algebra
$$\{ \Gamma_a (x), \Gamma_b (x^{\prime})
\} = i \delta_{ab} \delta (x - x^{\prime} ), $$
$$\{ \Theta_a (x), \Theta_b (x^{\prime})
\} = -i \delta_{ab} \delta (x - x^{\prime} ). \eqno(19)$$
where $\{\,\,,\,\,\}$ denotes symmetric Poisson bracket.

The next step is to impose that the \lq square root\rq $\,$  of the
bosonic theory is the fermionic theory (analogous to the
relation between the Klein-Gordon and Dirac equations), i.e.
$$\{ {\cal S}_a (x) , {\cal S}_b (x^{\prime}) \} = i \delta_{ab} {\tilde
L_b}(x) \delta (x - x^\prime). \eqno(20)$$

Using (18) and (19), we find that $\tilde L_a$ is
$$\tilde { L_a} = \alpha^2 h^2_a + \alpha\beta Q h^{\prime}_a - \alpha^2 i
{\Gamma}_a \Gamma^{\prime}_a - \gamma^2 J^2_{-a} - \gamma \delta Q
J^{\prime}_{-a} + \gamma^2 i \Theta_{-a} \Theta^{\prime}_{-a}. \eqno(21)$$

Which is the bosonic constraint modified by the presence of fermions.
If we now take the limit $(\Gamma, \Theta) \to 0$,
the coefficients $\alpha, \beta,
\gamma$ and $\delta$ can be explicitly computed
$$\alpha = 1= \gamma , \,\,\,\
\beta = {1\over {\sqrt{2}}}= \delta. \eqno(22)$$

Of course, our task is still incomplete because we have to show that the new
set of constraints $(\tilde{ L_a}, {\cal S}_a)$ satisfy a closed superalgebra.

We find
$$\eqalignno{&\{ {\cal S}_a (x) , {\cal S}_b (x^{\prime}) \} =  i \delta_{ab}
{\tilde L}_a (x ) \delta (x - x^{\prime}),
\cr &[ {\tilde L}_a (x), {\cal S}_b (x^{\prime}) ] = \delta_{ab}
( 2{\cal S}_a (x) + {\cal S}_a (x^{\prime}) ) \delta^\prime
(x - x^{\prime}),
\cr & [{\tilde L}_a (x) , {\tilde L}_b (x^{\prime}) ] = \delta_{ab}
({\tilde L}_a (x) +
 {\tilde L}_a (x^{\prime})) \delta^\prime (x  - x^{\prime}).&(23)\cr} $$

In consequence all the constraints are first class and the
canonical hamiltonian
$$H = \int dx \biggl( N^a \tilde{L}_a + i \lambda^a {\cal S}_a \biggr),
\eqno(24)$$
vanishes as a consequence of the reparametrization invariance generated by
$\tilde{L_a}$.

The hamiltonian action for this system is
$$\eqalignno{ S = \int d^2x &\biggl[ -{\tilde P}_0{\dot
{\tilde e}_0} + {\tilde P}_1{\dot {\tilde e}_1} -
{i\over 2}{\dot \Gamma}_a \Gamma_a- {i\over 2}{\dot \Theta}_a \Theta_a -
N^a{\tilde L_a} - i\lambda^a {\cal S}_a \bigg]  \cr & +
 Surface \,\,\, Terms. &(25)\cr}$$

Once the fermionic constraints are determined it is straightforward to compute
 the
local supersymmetry transformations that leave invariant the action (20)
$$\eqalignno{&\delta {\tilde e}_0 = -{{\sqrt 2}\over Q}( \xi_+ \Gamma_+
+ \xi_- \Gamma_-) ,
\cr & \delta {\tilde P}_0 = {Q\over {\sqrt 2}}\biggl[ {(\xi_+ \Gamma_+)}^\prime
- {(\xi_- \Gamma_-)}^\prime -{(\xi_+ \Theta_-)}^\prime +
{(\xi_- \Theta_+)}^\prime \biggr],
 \cr &\delta \Gamma_\pm = i\xi_\pm h_\pm - iQ \xi_\pm^\prime,
\cr &\delta {\tilde e}_1 = {{\sqrt{ 2}}\over Q}\biggl[ \xi_+ \Theta_- +
\xi_- \Theta_+\biggr],  \cr &\delta {\tilde P}_1 = {Q\over {\sqrt 2}}
 [{(\xi_+\Gamma_+)}^\prime +  {(\xi_+\Theta_-)}^\prime +
{(\xi_-\Gamma_-)}^\prime + {(\xi_-\Theta_+)}^\prime ],
\cr &\delta \Theta_\pm = -i [ \xi_\mp J_\mp -
Q\xi^\prime_\mp], &(26) \cr}$$
plus appropiate transformation rules for the lagrange multipliers
$N_a$ and $\lambda_a$ that are interpreted as the graviton and gravitino
respectively.

It is necessary to be cautious with the surface
term that appears in (24). The presence of this term implies the
invariance under local supersymmetry if and only if at the end points the
local supersymmetry parameters satisfy
$$\xi_+ (t_1, x) + \xi_- (t_2, x) = 0, $$
$$\xi_+ (t_1, x) - \xi_- (t_2, x) = 0. \eqno(27) $$

Now we discuss $2D$ (super)gravity with string-type boundary conditions,
i.e., the space direction is compact, of course with suitable
boundary conditions so that the boundary terms in (6) vanish.
In this case, in analogy to string thery, we can expand the constrints
in Fourier Series.

For the bosonic case, with periodic boundary
conditions, the constraints are now:
$$\eqalignno{ L_n  &=  {1\over 2\pi} \int_{-\pi}^{+\pi} dx e^{inx} L_+
(x) =  \cr &= \sum_{-\infty}^{+\infty} \alpha_{n-m} \alpha_m -
\sum_{-\infty}^{+\infty} \beta_{n-m} \beta_m -
{inQ\over 2{\sqrt \pi}} (\alpha_n-\beta_n),  &(28)\cr}$$
where the time dependent oscillator variables
$\alpha_n$ and $\beta_n$ are defined as
$$\alpha_n = {1\over \sqrt{2\pi}} \int_{-\pi}^{+\pi} dx e^{inx} h_+(x),$$
$$\beta_n = {1\over \sqrt{2\pi}} \int_{-\pi}^{+\pi} dx e^{inx} J_-(x),
\eqno(29)$$
and satisfy the equal time Poisson commutators
$$\lbrack \alpha_n , \alpha_m \rbrack = -2i n \delta_{n+m},$$
$$\lbrack \beta_n , \beta_m \rbrack = 2i n \delta_{n+m}.\eqno(30)$$

Using (28)-(30), we find that the classical Virasoro algebra is
$$\lbrack L_n , L_m \rbrack = - i(n-m) L_{n+m}. \eqno(31)$$

Canonical quantization is now straightforward, equal-time Poisson
brackets are substituted by equal-time commutators. The computation
of the central charge is as usual and it can be easily seen that
the contribution of the modes $\beta_n$ cancel the one of the
$\alpha_n$ so, that the resulting central charge is zero.
In the supergravity case we have to distinguish between the Ramond and
Neveu-Schwarz sectors, thus the constraints for open boundary conditions are
$$\eqalignno{ L_n &= \sum_{-\infty}^{+\infty} \alpha_{n-m} \alpha_m -
\sum_{-\infty}^{+\infty} \beta_{n-m} \beta_m - {inQ\over 2\sqrt{\pi}} (\alpha_n
 -
 \beta_n)
\cr & + \sum_{-\infty}^{+\infty} m \gamma_{n-m}\gamma_m -
 \sum_{-\infty}^{+\infty} m d_{n-m} d_m ,&(32) \cr}$$
and
$$S_n = \sum_{-\infty}^{+\infty}
\gamma_{n-m}\alpha_m - {iQn\over 2\sqrt{\pi}}\gamma_n
+ \sum_{-\infty}^{+\infty}
d_{n-m}\beta_m - {iQn\over 2{\sqrt{\pi}}}d_n, \eqno(33)$$
for the Ramond sector. Here
$$\gamma_n = {1\over \sqrt{2\pi}} \int_{-\pi}^{+\pi} dx e^{inx} \Gamma_+ (x)$$
$$d_n = {1\over \sqrt{2\pi}} \int_{-\pi}^{+\pi} dx e^{inx} \Theta_-(x),
\eqno(34)$$
are the fermionic oscillator variables that satisfy the symmetric Poisson
 algebra
$$\eqalignno{ &\lbrace \gamma_n , \gamma_m \rbrace = i  \delta_{n+m}
\cr &\lbrace d_n ,d_m \rbrace = i \delta_{n+m}.&(35)\cr}$$

In the Neveu-Schwarz sector the constraints are
$$\eqalignno{ L_n &= \sum_{-\infty}^{+\infty} \alpha_{n-m} \alpha_m -
\sum_{-\infty}^{+\infty} \beta_{n-m} \beta_m - {inQ\over 2\sqrt{\pi}} (\alpha_n
 -
 \beta_n)
\cr & + \sum_{r\in Z+{1\over 2}} r \gamma_{n-r}\gamma_r -
 \sum_{r\in Z+{1\over 2}} r d_{n-r} d_r .&(36) \cr}$$

and
$$F_n = \sum_{-\infty}^{+\infty}
\gamma_{r-m}\alpha_m - {iQn\over 2\sqrt{\pi}}\gamma_r
+\sum_{-\infty}^{+\infty}
d_{r-m}\beta_m - {iQr\over 2\sqrt{\pi}} d_r. \eqno (37)$$

In both sectors the Virasoro algebras are (R)
$$\eqalignno{ &\lbrack L_n , L_m \rbrack = - i(n-m) L_{n+m},
\cr & \lbrack L_n, S_m \rbrack = -i(n-2m) S_{n+m}, \cr &
\lbrace S_n , S_m \rbrace = i L_{n+m}, &(38) \cr}$$
and (NS)
$$\eqalignno{ &\lbrack L_n , L_m \rbrack = - i(n-m) L_{n+m},
\cr & \lbrack L_n, F_r \rbrack = -i(n-2r) F_{n+r}, \cr &
\lbrace F_r , F_s \rbrace = i F_{r+s}, &(39) \cr}$$

Hier the situation after quantization is the same as in the
bosonic case, the central charge contributions of all modes
cancel to give zero.

Therefore, in conclusion, taking into account the fact that the
conformal anomaly does not depend on the gauge fixing, our
result means that the theory does not have physical degrees
of freedom. In other words, if we fix the conformal gauge
the only contribution to the total central charge will be the
one of the ghost fields meaning that our theory describes only
gauge degrees of freedom. Further, our results depend on the
fact that we chose a manifold wiht trivial topology, a stripe
or a cylinder. However, if nontrivial boundary conditions are
chosen, a nontrivial physical content for the quantum theory
can arise as shown by Marnelius {\bf [9]}. In this context,
the fact that our formalism is gauge covariant allows a
bigger freedom in the choice of nontrivial topologies
(boundary conditions). Work is in progress in this direction.

{\bf Acknowlegdments}. We would like to thank M. Asorey, J.L. Cort\'es,
H.B. Nielsen, S. Ouvry and G. Sierra for the discussions. J.G. thanks
A. Macias and H. Salazar at the UAM and UAP (M\'exico) for their hospitality.
C.R. Thanks J.F. Cari\~nena at the Departamento de Fis\'{\i}ca
Te\'orica, Zaragoza for his hospitality. J.G. was partially supported by
FONDECYT Grant 0867/91 and CNRS. C.R was partially
supported by CONACYT-M\'exico Grant 0758-E9109.
\vskip 0.10cm
\centerline{\bf References}
\vskip 0.15cm
\item{\bf {[1]}} A.M. Polyakov,{\it Phys. Lett.} {\bf B103}(1981)207.
\item{\bf {[2]}} A.M. Polyakov, {\it Mod. Phys. Lett.} {\bf A2}(1987)893.
\item{\bf {[3]}} V.H. Knizhnik, A.M. Polyakov and A.B. Zamolodchikov, {\it Mod.
 Phys.
Lett.} {\bf A3}\break (1988)819.
\item{\bf {[4]}} F. David, {\it Mod. Phys. Lett.} {\bf A3}(1988)1651;
J. Distler and T. Kawai, {\it Nucl. Phys.} {\bf B321}(1989)509.
\item{\bf {[5]}} D.V. Boulotov, V.A. Kazakov, I.K. Kostov and A.A. Migdal,
{\it Nucl. Phys.} {\bf 275}(1987)\break 218; F. David, J. Jurkiewicz, A.
 Krzywicki and
B. Petersson,{\it Nucl. Phys.} {\bf B290}(1987)\hfill \break 218.
\item{\bf {[6]}} D.J. Gross and A.A. Migdal, {\it Phys. Rev. Lett.} {\bf
 64}(1990)717.
\item{\bf {[7]}} E.S. Egorian and R.P. Manvelian, {\it Mod. Phys. Lett.} {\bf
A}(1991)2371.
\item{\bf {[8]}} E. Abdalla, M.C.B. Abdalla
, J. Gamboa and A. Zadra,{\it Phys. Lett.} {\bf B273}(1991)222.
\item{\bf {[9]}} R. Marnelius,{\it Nucl. Phys.} {\bf 211}(1983)14 and
{\it ibid}
 {\bf 221}(1983)337.
\item{\bf {[10]}} T. Regge and C. Teitelboim, {\it Ann. of Phys. (N.Y.)} {\bf
88}(1974)286.
\item{\bf {[11]}} P.A.M. Dirac,{\it Proc. Royal. Soc.} {\bf 38}(1928)610.
\item{\bf {[12]}} C. Teitelboim,{\it Phys. Rev. Lett.} {\bf 38}(1977)1106.
\item{\bf {[13]}} M. Henneaux and C. Teitelboim,{\it Ann. of Phys.} (N.Y){\bf
 143}
(1982)127 .
\item{\bf {[14]}}  For a discussion on the surface terms in the spinning
string case see, L. Mezincescu and R.I. Nepomechie, {\it Nucl. Phys.} {\bf
B322}(1989)127.
\end